\documentstyle[aps,twocolumn]{revtex}
\tightenlines

\begin{document}
\title{Local magnetic structures induced by inhomogeneities of the lattice
in $S=1/2$ bond-alternating chains and response to time-dependent magnetic
field with a 
random noise}

\author{Masamichi Nishino}

\address{Department of Chemistry, Graduate School of Science,\\
Osaka University, Toyonaka, Osaka 560, Japan\\} %

\author{Hiroaki Onishi}
\address{Department of Earth and Space Science, Graduate School of Science,\\
Osaka University,       Toyonaka, Osaka 560, Japan}%

\author{Kizashi Yamaguchi}

\address{Department of Chemistry, Graduate School of Science,\\
Osaka University, Toyonaka, Osaka 560, Japan\\} %

\author{Seiji Miyashita}
\address{Department of Applied Physics, University of Tokyo\\
Bunkyoku, Tokyo, Japan}

\maketitle

\begin{abstract}

We study effect of inhomogeneities of the lattice in the $S=1/2$
bond-alternating chain by using a quantum Monte Carlo method and an exact
diagonalization method.
We adopt a defect in the alternating order as the inhomogeneity and we call it 
{\lq}{\lq}bond impurity{\rq}{\rq}. 
Local magnetic structures induced by the bond impurities
are investigated both in the ground state and at very low temperatures.
The local magnetic structure can be looked on as an effective $S=1/2$ spin
and the weakness of the interaction between the local structures causes the
quasi-degenerate states in the low energy.
We also investigate the force acting between bond impurities and 
find that the force is generally attractive.
We also study the dynamical property of the local magnetic structure.  
While the local magnetic structure behaves as an isolated $S=1/2$ spin in the 
response to a time-dependent uniform field, it is found to be robust
against the effect of a random noise applied at each site individually in
the sweeping filed.

\end{abstract}

\section{Introduction}

In the low dimensional quantum spin systems, it has been turned out that 
the quantum mechanical 
effect, in particular the mechanism of the singlet pair formation, 
plays an important role for the ground state spin configuration~\cite{Haldane,MG,AKLT}.
There have been found a wide variety of peculiar arrangements of the 
singlet pairs in the ground state which brings new types of singlet 
(or nonmagnetic) ground state phases such as VBS~\cite{AKLT}, RVB~\cite{Anderson,Fazekas}, etc.
As general characteristics of these systems, 
a finite energy gap exists between the ground
state and excitation states and the correlation length in the ground state is
finite.  If a lattice has some inhomogeneities such as impurity sites, 
defects of periodicity, and edge points, etc., then local magnetic structures are induced around them~\cite{Sorensen,Pascal,Kennedy,Miyashita}.
Property of such induced moments is one of the most interesting
current topics. 

In this paper we will study the interaction between the
induced moments and also dynamical properties of such moments in the  
bond-alternating Heisenberg antiferromagnetic (HAF) chain.
This model is one of the simplest systems of the singlet ground state
which consists of singlet states at strong bonds. 
The magnetic susceptibility and specific-heat of this model 
was studied in detail as a function of the ratio of the alternation coupling
constants by Duffy and Barr~\cite{Duffy_theory}.
Various materials corresponding to this model have been studied experimentally,
i.e., aromatic free-radical compounds~\cite{Duffy_exp},
Cu(NO$_3$)$_2\cdot$2.5H$_2$O~\cite{Diedrix}, and
(VO)$_2$P$_2$O$_7$~\cite{Garrett}, etc.
The ratio of the alternation coupling constants for theses materials 
has been estimated comparing with the theoretical results~\cite{Duffy_exp}.
This system has been also extensively studied  concerning with the
spin-Peirels transitions~\cite{Bray,Uchinokura,Nishi}. 

Here we consider a defect of the alternating order of the bonds, such as
$\cdots$ABAB\underline{AA}BAB$\cdots$ or $\cdots$ABA\underline{BB}ABAB$\cdots$,
where A or B denotes the strength of bonds (the magnitude of exchange constant). 
At this defect the singlet dimer state cannot be formed there.
First we study how a magnetic structure appears at these positions. 
In this paper we call such a defect of the alternating order 
{\lq}{\lq}bond impurity{\rq}{\rq}. 
Furthermore, the induced magnetic structure depends on whether 
the edge bond is A or B.
Bond impurity effects in the uniform HAF have been investigated 
in our previous paper~\cite{Nishino} and we compare the role of 
the bond impurity in the present model with them.

Next, we focus on what kind of interactions act between these locally induced
magnetic structures. We study the force by
investigating the dependence of the ground state energy on the distance
between the impurities for various types of configurations of defects. 
The force between the bond impurities has been  also studied for the
uniform HAF~\cite{Nishino}. There we found that the attractive force
acts.  In the present model we again find only attractive force 
regardless of the types of configuration.

Due to the finite correlation length of the present model, the induced 
moments behave almost independently.
Thus we can regard such a local moment as a microscopic moment.
Because recently the technology in microscopic processing makes remarkable
progress and the analyses in microscale or nanoscale phenomena have become
possible, the quantum phenomena in microscale or nanoscale have received
much attention. For example, the resonant tunneling phenomena 
have been observed in recent experiments on high-spin molecules
(Mn$_{12}$)~\cite{Friedman,Thomas,Hernandez} and the concept of quantum
tunneling of the magnetization (QTM) has become a topic of interest.  
Recently the adiabatic motion of spin $1/2$ was observed in V$_{15}$
and the effects of thermal disturbance on the dynamics has been discussed~\cite{Chiorescu}.
We have studied the tunneling dynamics from viewpoint of the nonadiabatic
transition~\cite{Miya1,DeRaedt,Miya2,TFE}.

Under a time-dependent magnetic field, how does that local magnetic structure 
behaveH We study the response of the
magnetization to a sweeping magnetic field. 
Although the induced moment consists of several spins, the total
magnetization behaves as a single spin when a uniform field is applied.
Under a sweeping magnetic field the behavior of magnetization is described
by the Landau-Zener-St$\rm{\ddot{u}}$ckelberg (LZS)
formula~\cite{Landau,Zener,Stuckel}.  
In realistic situations, noise disturbs the simple behavior of LZS 
formula~\cite{Kaya1,Kaya2}.
Furthermore, if the noise acts independently at each site, the behavior
of induced moment (a cluster of spins) is expected to be different from
that of isolated single spin. We investigate effects of such individual 
noise at each site on the dynamical properties.

This paper is organized as follows.
In the next section, we explain briefly the method used in this study.
In Sect. \ref{alternate}, effects of bond impurities on the magnetic 
structure in bond-alternating chains are studied.  
In Sect. \ref{force}, we study the force between bond impurities.
In Sect. \ref{response}, we investigate the response of the magnetization
to a sweeping field.
Sect. \ref{sec:summary} is devoted to the summary and discussion.

\section{Model and Method}
The Hamiltonian treated in the present paper is given by
\begin{equation}
{\cal H}=\sum_{i} J_{i,i+1}{\bf S}_{i}\cdot{\bf S}_{i+1}, 
\label{model}
\end{equation}
where ${\bf S}_{i}=(S_i^x, S_i^y, S_i^z)$ are the $S=1/2$ spin operators at
the site $i$.
We study the bond-alternating chain where $J_{i,i+1}$ changes alternately 
among $J_1$ (a strong bond) and $J_2$ (a weak bond).
Here we consider defects of the alternation which cause
inhomogeneities of the lattice (bond impurities).
To study low temperature properties of the model, 
we mainly use the loop algorithm of the continuous time quantum 
Monte Carlo method (LCQMC)~\cite{loop,cont} with a method of
specification of the magnetization $M_z$~\cite{Pascal,Nishino}.
This method overcomes the problem of long autocorrelation in Monte 
Carlo update and allows us to study systems at very low
temperatures.  
In the present work, we performed $10^5$ Monte Carlo steps (MCS) 
for getting equilibrium of the system and $10^6$ MCS to obtain quantities
in the equilibrium state.  Here a MCS means an update of the whole 
spins.

In order to study the dynamical response of spins under 
a time-dependent external field, we investigate the following system:
\begin{equation}
{\cal H}=\sum_{i} J_{i,i+1}{\bf S}_{i}\cdot{\bf S}_{i+1}+
2\Gamma\sum_{i}S_i^x-\sum_{i}(H(t)+h_i(t))S_i^z, 
\label{eq_noise}
\end{equation}
where 
$H(t)$ is a time-dependent external 
field and $h_i(t)$ is a random noise applied to each site individually.
$\Gamma$ is the transverse field which represents terms for 
quantum fluctuation of $M_z$. 
The time evolution of the state is obtained by the time-dependent
Schr$\rm{\ddot{o}}$dinger equation 
(TDSE)
\begin{equation}
i\hbar\frac{\partial}{\partial t}|\Psi (t) \rangle={\cal H}|\Psi (t)\rangle, 
\label{eq:sch}
\end{equation}
where $|\Psi (t)\rangle$ denotes the wave function of the spin system at
time $t$.
We set $\hbar=1$.
Equation (\ref{eq:sch}) is solved using the fourth-order fractal
decomposition~\cite{Suzuki}, 
\begin{equation}
e^{-it \cal{H}}=[S_2(-it p_2)]^2 S_2(-it (1-4 p_2)) [S_2(-it p_2)]^2, 
\end{equation}
where $S_2(x)=e^{x {\cal H}_1/2}e^{x {\cal H}_2}e^{x {\cal H}_1/2}$ 
with $p_2=(4-4^{\frac{1}{3}})^{-1}$,
putting
\begin{eqnarray*} 
{\cal H}_1&=&\sum_{i} J_{i,i+1}{\bf S}_{i}\cdot{\bf S}_{i+1}+
2\Gamma\sum_{i}S_i^x, \\
{\cal H}_2&=&-\sum_{i}(H(t)+h_i(t))S_i^z.
\end{eqnarray*}

As the initial state, 
we set the applied field to its minimum value $H(t=0)=-H_0<0$, and put
the system to be the ground state for this field.
Then we sweep the filed as 
\begin{equation}
H(t)=-H_0+ct. 
\end{equation}
Beside this sweeping field, we provide a random noise
with an exponential-decaying autocorrelation function  
$\{h_i(t)\}$ by a Langevin equation
(Ornstein-Ulenbeck process),
\begin{equation}
\dot{h}(t)=-\gamma h(t) + \eta(t).
\label{Langevin}
\end{equation}
Here $\eta(t)$ is a white gaussian noise, 
\begin{equation}
\langle \eta(t) \rangle=0   \;\;\;{\rm and}\;\;\;
\langle \eta(0)\eta(t) \rangle=A^2\delta(t),
\label{white}
\end{equation}
where $A$ is amplitude, $\gamma$ is damping factor.
Thus obtained random process $h(t)$ has the properties
\begin{equation}
\langle h(t) \rangle=0 \;\;\;{\rm and}\;\;\;
\langle h(0)h(t) \rangle=\frac{A^2}{2\gamma}
{\rm exp}(-\frac{t}{\tau}), 
\label{rand_noise}
\end{equation}
where $\tau=1/\gamma$.
The dynamics of magnetization is obtained as 
\begin{equation}
M(t)=\langle \Psi (t)| \sum_i S_i^z |\Psi (t) \rangle.
\label{Eq_Mz}
\end{equation}
If the magnetic field changes rather slowly, i.e., the sweep rate is rather
small, $|\Psi (t)\rangle$ changes adiabatically. In this case the state
stays in the ground 
state of the system for the current field $H(t)$, 
and the magnetization follows 
the value of the ground state. When the sweeping rate becomes 
large the system cannot follow the change of the field completely,
and the nonadiabatic transition occurs. 
The probability to stay in the ground state was given by 
LZS formula as
\begin{equation}
p=1-{\rm exp}(\frac{-2\pi \Gamma^2}{c}).
\label{prob}
\end{equation}
When the effect of noise does not become negligible, 
the noise disturbs the quantum process and this probability
changes~\cite{Kaya1}.

\section{bond impurity in bond-alternating chains}
\label{alternate}

We investigate magnetic structures in the system (\ref{model}) with a
bond alternation 
$\cdots J_{1} J_{2} J_{1} J_{2}   \cdots$, where $J_{1} > J_{2}$. 
We study effects of a defect in the alternation, such as 
$\cdots  J_{1} J_{2} J_{1} J_{2}  \underline{J_{1} J_{1}} J_{2} J_{1} J_{2}
J_{1}  \cdots $.
Beside this defect, magnetic structures can be induced at the edges
as naturally understood from the VBS picture, i.e., an unpaired spin
causes an induced magnetization (see the figures).
Thus we study the following four systems of
the bond configurations:\\
(a) a chain of 63 sites where 
the two strong bonds are at the center and both edges terminate 
with a strong bond,\\
($J_{1} J_{2} \cdots J_{1} J_{2} \underline{ J_{1} J_{1}}
 J_{2} J_{1} \cdots J_{2} J_{1} $)\\
(b)  a chain of 65 sites where 
the two strong bonds are at the center and both edges terminate 
with a weak bond,\\
($J_{2} J_{1} \cdots J_{1} J_{2} \underline{ J_{1} J_{1}}
 J_{2} J_{1} \cdots J_{1} J_{2} $)\\
(c)  a chain of 63 sites where 
the two weak bonds are at the center and both edges terminate 
with a weak bond, \\
($J_{2} J_{1} \cdots J_{2} J_{1} \underline{ J_{2} J_{2}}
 J_{1} J_{2} \cdots J_{1} J_{2} $)\\
and \\
 (d)  a chain of 65 sites where 
the two weak bonds are at the center and the edges terminate 
with a strong bond\\
($J_{1} J_{2} \cdots J_{2} J_{1} \underline{ J_{2} J_{2}}
 J_{1} J_{2} \cdots J_{2} J_{1}$).\\
Here we take the strong bond to be $J_{1}=1.3$ and the weak bond to be
$J_{2}=0.7$. For this set of bonds,
the correlation length is estimated as $\xi \simeq0.82$ (see Appendix).
Here we take $k_{\rm B}$ as a unit of energy ($k_{\rm B}=1$).
These four models have odd number of spins and their ground state is
doublet according to the Lieb-Mattis theorem~\cite{Liep}.
We performed simulations at $T=0.01$ in the $M_z=1/2$ space to study
magnetic structures in the low energy state. 
The magnetization profiles $\{m_i\}$ of (a)-(d) are drawn in Fig.
\ref{fig_bondalt_mag}, 
where $m_i=\langle S_i^z\rangle$ and $\langle \; \rangle$ denotes the
canonical average at a given temperature.
A magnetization is induced locally around the impurity. 
First we consider the cases where the bonds at edges are strong, i.e., 
the model (a) and (d).
If we allocate a singlet pair at each strong bond, neighboring two strong bonds remain at the center in the model (a), while
one site remains in the model (d). The magnetization of $M_z=1/2$ is
assigned in the remaining part at the center of the lattices to induce a
local magnetic structure.
Because the edge bonds are strong, no magnetization is induced at the edges.  
Figures \ref{fig_bondalt_mag}(a) and (d) are considered to describe well the 
magnetization profiles of the ground state since these are gapful systems
without quasi-degenerate states within a subspace with a fixed
magnetization (i.e., $M_z=1/2$). 
We find that $M_z$ is distributed only into the $\pm1/2$ space in the simulations at this temperature.

In the models (b) and (c), 
when we allocate singlet state at the strong bond, there are three
positions for 
magnetic moments, i.e., the center and both edges.
Indeed in the both models (b) and (c), magnetizations are 
induced around the impurity and both edges as shown in Fig. 1. 
In order to study the distribution of magnetization in the lattice, we
introduce 
the summation of the magnetization per site from the left edge site
\begin{equation}
{\cal M}_z(j)=\sum_{i=1}^j m_i.
\label{tmag}
\end{equation}
We show this quantity for the model (b) in Fig. \ref{fig_bondalt_sum}. 
There the values of the left plateau and right plateau are 0.166 and 0.333,
respectively. 
From this figure we find a spin 1/6 locates at each local structure. This
deceptive fractional magnetization is considered to come
from mixing of states.

Looking on the local magnetic structure as an effective $S=1/2$ spin
interacting by an effective exchange $\tilde{J}$, 
this system is modeled by a 
three-site Heisenberg model ${\cal H}=\tilde{J}{\bf S}_{1}\cdot{\bf S}_{2}+
\tilde{J}{\bf S}_{2}\cdot{\bf S}_{3}. $
In the $M_z=1/2$ space the eigenvalues are 
\begin{equation}
E_{1}=-\tilde{J},  E_{2}=0$, and $E_{3}=\tilde{J}/2.
\label{ground eq}
\end{equation}
The corresponding eigenvectors are denoted by $|\phi_{i}\rangle, (i=1,2, 
{\rm and} 3)$.
The expectation values of magnetization of spins in each state are
\begin{eqnarray}
&& \langle \phi_{1}|S_{1}^z|\phi_{1}\rangle=\langle \phi_{1}|
S_{3}^z| \phi_{1}\rangle=1/3 \;\; {\rm and} \;\;  \langle
\phi_{1}|S_{2}^z|\phi_{1}\rangle=-1/6, \nonumber \\
&& \langle \phi_{2}|S_{1}^z| \phi_{2}\rangle=\langle \phi_{2}|S_{3}^z| 
\phi_{2}\rangle=0  \:\:\;\;\;\; {\rm and} \;\; \langle
\phi_{2}|S_{2}^z|\phi_{2}\rangle=1/2, \\
&& \langle \phi_{3}|S_{1}^z| \phi_{3}\rangle=\langle \phi_{3}|
S_{2}^z| \phi_{3}\rangle=\langle \phi_{3}|S_{3}^z| \phi_{3}\rangle=1/
6. \nonumber  
\end{eqnarray}
In the gapped spin system the correlation function decays exponentially
and the effective coupling between the induced moments is expected to be
very small\cite{Pascal,Kennedy}, i.e., $\tilde{J} \ll1$.
Thus these states are considered to
be almost degenerate even at this temperature ($T=0.01$),
although there are energy gaps of order $O(\tilde{J})$ 
between the state $|\phi_{1}\rangle$, 
$|\phi_{2}\rangle$, and $|\phi_{3}\rangle$. 
In such a case the three states appear in the equal
probability, and the expectation value of magnetization is given by an equal-weight average in the three state. That is,  $\langle S_{1}^z \rangle$, $\langle S_{2}^z
\rangle$, and 
$\langle S_{3}^z \rangle$ are given 
by $(1/3+0+1/6)/3=1/6$, $(-1/6+1/2+1/6)/3=1/6$, and $(1/3+0+1/6)/3=1/6$,
respectively. 
These values correspond to the observed deceptive fractional magnetization. 

In order to  confirm the above modeling we perform the following two
investigations.  
First, we investigate a short chain of $L=21$ by an exact diagonalization method
in order to check that the ground state of the type (b) is represented 
by $| \phi_{1}\rangle$ of the three-spin model.
Here we choose a chain of a shorter 
correlation length because the length of the chain is short. 
Namely, we set $J_{1}=2$ and $J_{2}=0.5$, where $\xi \ll 1$.
Fig. \ref{fig_bondalt_diagmag} shows the
summation of the magnetization from the left edge site (Eq. (\ref{tmag}))
for this model.
The local magnetization around the left edge is about 1/3 which corresponds to $\langle \phi_{1}|S_{1}^z| \phi_{1}\rangle$. This causes the left plateau.
The local magnetization around the impurity at the center is about $-$1/6 ($\langle
\phi_{1}|S_{2}^z|\phi_{1}\rangle$). Therefore the right plateau is 1/6 ($\langle \phi_{1}|S_{1}^z| \phi_{1}\rangle$ + $\langle\phi_{1}|S_{2}^z|\phi_{1}\rangle$). 
Finally adding the local magnetization around the right edge (1/3), ${\cal M}_z(21)$ 
terminates at 1/2.
Thus the ground state $| \phi_{1}\rangle$ represents well that of
the type (b) model.

Second, to confirm the quasi-degeneracy in the model (b), we check the 
 distribution of $M_z$ in the Monte Carlo simulation, which is shown in
Fig. \ref{fig_bondalt_prob}. 
Here the distribution for $M_z=3/2$ is about 12.5 $\%$.  
Because there are three states in the $M_z=1/2$ space and one state in the
$M_z=3/2$ space in the three-spin model, this distribution indicates that
these four states are equally populated and that $T=0.01$ is much higher
than the energy gaps between these four states.  
In the three-spin model the eigenvalue of the state of $M_z=3/2$ is
$\tilde{J}/2$, while the eigenvalues of the state of $M_z=1/2$ are
$-\tilde{J}$, 0, and $\tilde{J}/2$ (Eq. (\ref{ground eq})).
In the $M_z=3/2$ space we observe  $S=1/2$ moment at each local structure 
which represents the state 
$|+++\rangle$ in the effective three-site model. The values $\{m_i\}$ are
almost three times as large as those of Fig. \ref{fig_bondalt_mag} (b).
In principle we can obtain the energy gap from the temperature 
dependence of the distribution~\cite{Pascal}.  
However it is too small to detect here.

We find a similar scenario for the model (c).
Thus we conclude that in bond-alternating systems, 
local magnetic structures are induced by a bond impurity or weak edge 
bonds as an effective $S=1/2$ spin and they behave almost
independently.

Now let us examine more detailed structures of the local magnetic structures.
 In the case (a) a negative magnetization appears at the middle site, while
in the case (d) a positive magnetization appears there.
The interaction of the three spins at the center of the model (a) is 
approximately represented by the three-spin model with the strong bonds. 
Here the magnetizations at the center of the model (a) are distributed as
about (1/3, $-1/6$,1/3). 
On the other hand in the model (d) a spin at the center is isolated from
the others.


The local magnetic structures in the present model are well isolated. Therefore 
we can locate such magnetic structures as we desire.
In Fig. \ref{five_moment}  we show a magnetization profile which
has five positions of induced structures in a chain of $L=101$.
This system has three $J_2J_2$ defects and weak edge bonds.
This configuration is obtained at a very low temperature ($T=0.01$) in the space of
the total magnetization $M_z=1/2$. Each moment almost behaves
independently. Thus magnetization at each location 
is given by an average over all the possible states. 
In the present case, each local moment has a deceptive magnetization $1/10$. 
Furthermore we confirmed that the distribution of magnetization $P(M_z)$ is
given 
by a binomial distribution 
\begin{equation}
P(M_z) = \frac{1}{2^5} \frac{5\: !}{(5-2|M_z|) \: ! \: (2|M_z|)\: ! }.
\end{equation}


\section{Force between two defects in an alternating chain}
\label{force}

When a bond-alternating system has impurities, what kind of interaction
does exist between them, attractive or repulsive? We study the force between the bond impurities in this section.
We estimate the ground state energies of the spin system in various fixed
configurations of bonds and 
compare the ground state energies as a function of the distance between the
bond impurities.  

In the alternating chain ($ \cdots J_{1} J_{2} J_{1} J_{2}  \cdots$ ), 
the system may have a pair of defects by shifting a position 
of a strong  bond by one. 
\begin{equation}
\cdots J_{1} J_{2} \underline{J_{1} J_{1} J_{2} J_{2} } J_{1} J_{2} 
\cdots .
\label{ichi}
\end{equation}
If we shift the position furthermore, the system has a configuration 
\begin{equation}
\cdots \underline{J_{1} J_{1}} J_{2} J_{1} \underline{ J_{2} J_{2} } 
\cdots , 
\label{ni}
\end{equation}
etc. We study dependence of the energy on the distance ($\Delta_a$) between
the positions of $J_{1} J_{1}$ and $J_{2} J_{2}$. 
 We define $\Delta_a=0$ for the case of no defect, $\Delta_a=1$ for the
configuration (\ref{ichi}),  $\Delta_a=2$ for the configuration (\ref{ni})
and so on.  
In Fig. \ref{fig_force_alt_diag} (a)
we plot the ground state energy as a function of $\Delta_a$
obtained by an exact diagonalization for $L=24$ with $J_1=2$ and 
$J_2=1$ in the periodic boundary condition (PBC). 
The ground state energy becomes larger as $\Delta_a$ becomes larger.
Therefore we find an attractive force between the impurities ($J_{1} J_{1}$
and $J_{2} J_{2}$).

Next another situation is considered.
If one $J_2$ is exchanged by $J_1$ in the alternating chain, 
the system has a configuration
\begin{equation}
\cdots J_{1} J_{2} \underline{J_{1} J_{1} J_{1}} J_{2} J_{1} 
\cdots .
\label{stand}
\end{equation}
We define $\Delta_b= 0$ for this configuration.
 Shifting a position of a $J_{1}$ by one, the system has 
two $J_{1} J_{1}$ pairs and has a configuration, 
\begin{equation}
\cdots\underline{J_{1} J_{1} } J_{2} \underline{J_{1} J_{1} }
\cdots .
\label{one_shf}
\end{equation}
We define this distance between $J_{1} J_{1}$ and $J_{1} J_{1}$
as $\Delta_b= 1$.
Furthermore, $\Delta_b=2$ is defined for the configuration (\ref{two_shf})
and etc.
\begin{equation}
\cdots\underline{J_{1} J_{1} } J_{2} J_{1} J_{2}\underline{J_{1} J_{1} }
\cdots .
\label{two_shf}
\end{equation}
The ground state energies for a PBC chain of $L=24$ are shown as a function of $\Delta_b$ in 
Fig. \ref{fig_force_alt_diag} (b). 
We also study the interaction 
between $J_{2} J_{2}$ and $J_{2} J_{2}$, where we define the distance between 
the impurities in the same way.
We again found 
that the ground state energy increases as the distance  becomes larger.
Thus it has been found that 
an attractive force acts between the impurities regardless of their type.

If we allow positions of impurities to move, the distance between
impurities would be distributed in the canonical distribution 
according to the interaction between the impurities in the thermal equilibrium 
at a given temperature as has demonstrated in the previous paper~\cite{Nishino}.

\section{responce of the local magnetic structure to the dynamical field}
\label{response}

In this section we investigate dynamical response of the
local magnetic structure (effective $S=1/2$ spin) induced by a bond
impurity to a time-dependent magnetic field with a random noise
in the dynamical model (\ref{eq_noise}).

First let us consider the response of the magnetization of a free $S=1/2$ spin to 
a sweeping field.
The Hamiltonian is given by
\begin{equation}
{\cal H}(t)=2\Gamma S_x-(-H_0+ct) S_z.
\label{free_spin}
\end{equation}
This system is two-level system whose energy levels are shown in
Fig.~\ref{level_cross} as a function of $H$, where the avoided level
crossing occurs near $H=0$.
If $H_0 \gg \Gamma >0$, the ground state consists primarily of the
$M_z=-1/2$ state at $t=0$.
The probability for the system to end up in the 
$M_z=1/2$ state (i.e., the probability to change its magnetization) at
$t=\infty$ is 
given by Eq. (\ref{prob})

Next we consider the case of a local magnetic structure (effective $S=1/2$)
on a lattice.
When the noise $\{h_i(t)\}$ does not exist, dynamics of the system is
generally the same as that of a free $S=1/2$ spin as far as we concern the
states in an $S=1/2$ space. 
It is easily understood as follows.
Adopting $\{ |+\rangle, |-\rangle\}$ as the basis set, the matrix
representation of Eq. (\ref{free_spin}) is 
\begin{equation}
{\cal H}  = \pmatrix{
                    -\frac{H}{2}  &  \Gamma \cr
                     \Gamma        &  \frac{H}{2} \cr
                        }.
\end{equation}
Noting that $\sum_{i} J_{i,i+1}{\bf S}_{i}\cdot{\bf S}_{i+1}$ and
$2\Gamma\sum_{i}S_i^x-H(t)\sum_{i}S_i^z$ commute, the matrix representation
of Eq. (\ref{eq_noise}) in the total $S=1/2$ space is given by
\begin{equation}
{\cal H}  = \pmatrix{
                    -\frac{H}{2}+{\rm const.} &  \Gamma \cr
                     \Gamma                 &  \frac{H}{2}+{\rm const.} \cr 
                        },
\label{lowest_two}
\end{equation}
adopting $\{ |S^{\rm tot}=1/2, M_z=1/2 \rangle, |S^{\rm tot}=1/2,M_z=-1/2
\rangle\}$ as the basis set.
The constant in Eq. (\ref{lowest_two}) does not depend on $H(t)$. 
Thus the dynamics is independent of the number of sites and the combination
of $\{J_{ij}\}$.

In the experimental situation, however, noise usually has an influence on
the system. 
If a random noise is applied to each site individually, the term
$-\sum_{i}(H(t)+h_i(t))S_i^z$ does not commute with $({\bf S}^{\rm
tot})^2=(\sum {\bf S}_i)^2$, and therefore the dynamics would be changed
when the interaction $\{J_{ij}\}$ and the system size are varied.

Here for the study of dynamical properties we adopt the minimum model of the type (d) because we can treat a limited number of spins in the scheme of Eq.
(\ref{eq:sch}). This model consists of five sites and we take $J_1=1.0$ and
$J_2=0.5$.  This is the minimum bond-alternating model with a
$J_2J_2$ bond impurity. 
This system is looked on as an effective $S=1/2$ spin. 

It would be expected that
the effect of noise is reduced as the system size becomes large and 
a local magnetic structure (effective $S=1/2$ spin) consisting of several
number of spins is less sensitive to such a random noise 
than a free $S=1/2$ spin.
In order to study effects of the noise we investigate the following properties.

First we study the broadening of the level due to the noise. 
A energy gap and a sweeping rate at the level crossing point are important
factors to determine the transition probability under a sweeping uniform field
(see Eq. (\ref{prob})).
It would be useful to investigate how the noise influences the energy gap at the
level crossing point.
For a single spin the energy gap is given by
\begin{equation}
\Delta E=\sqrt{(2\Gamma)^2 \; + \; (h(t))^2}
\end{equation}
for $H(t)=0$.
Thus using the distribution of $h(t)$ (a gaussian distribution with the
variance 
$A^2/2\gamma$ in Eq. (\ref{rand_noise})), we can calculate the mean
$\langle \Delta E \rangle$ and the width $\delta E=\sqrt{ \langle (\Delta
E)^2 \rangle- \langle \Delta E \rangle^2}$, 
which are listed in Table \ref{Delta_E}.
For the local magnetic structure, we calculate the energy gap in a noise 
by a perturbation method. That gives 
\begin{eqnarray}
\Delta E &=&2\Gamma + \frac{2 |\langle \phi_1^{(0)} | V_1 | \phi_2^{(0)}
\rangle|^2}{E_2^{(0)}-E_1^{(0)}}    \\
&+& {\rm contributions \; from \; higher \; levels}, \nonumber
\end{eqnarray}
where $E_1^{(0)}$ and $\phi_1^{(0)}$ ($E_2^{(0)}$ and $\phi_2^{(0)}$) are the 
eigenstate of the ground state (the first exited state) of the non 
perturbed hamiltonian. The term $V_1$ is $\sum_i h_i S_i^z$.
Noting that 
\begin{equation}
\langle \; |\langle \phi_1^{(0)} | V_1 | \phi_2^{(0)} \rangle|^2 \rangle = 
\sum_i \langle h_i^2 \rangle |\langle \phi_1^{(0)} | S_i^z | \phi_2^{(0)}
\rangle |^2,
\end{equation}
we can calculate $\langle \Delta E \rangle$ from the 
distribution of $\{h_i\}$, which are also listed in Table \ref{Delta_E} for 
$L=5$ and $L=9$.
For large values of $A$, we obtain the energies by diagonalizing the 
hamiltonian Eq. (\ref{eq_noise}) and obtained $\langle \Delta E \rangle$
and $\delta E$ numerically from 500 samples of $\{h_i\}$, which are also
listed in Table \ref{Delta_E}.
Thus we find the noise causes almost the same effect on the broadening of the 
energy levels at $H(t)=0$ in both systems, i.e., a 
single spin and local magnetic structures, which is not in accordance with
the above expectation.

Second, we investigate the dynamical properties of the both systems. 
In particular, we study the time evolution of magnetization under the 
sweeping field.
We set parameters
$\Gamma=0.02$, $H_0=0.5$, $c=0.0005$, $t_{\rm max}$=2000, and 
$dt=0.01$.
In this parameter set the probability $p$ in Eq. (\ref{prob}) is nearly 1.
As was mentioned, the responses of the magnetization 
are the same in both systems when a random noise is not applied. 
Time evolution without noise is shown by thin dotted lines in Figs.
\ref{comp_mag} 
(a)-(c). 

Next we investigate the dynamics with the random field (Eq. (\ref{Langevin})).
We observe the two cases changing the amplitude $A$ of noise (Eq.
(\ref{white})); (a) $A=0.01$ and (b) $A=0.02$. The value of
$\gamma$ is fixed at 0.1.
The lowest two energy levels of the system with a
random noise $\{h_i(t)\}$ for case (b) are illustrated as a function of
time in Fig. \ref{level_noise}.

Figure \ref{comp_mag} (a) shows $M(t)$ in the noise of a small amplitude
$(A=0.01)$. 
The transition probabilities for 
 a local magnetic structure (effective $S=1/2$ spin) and an $S=1/2$ free 
spin are very close to each other at this small amplitude and the
probabilities are reduced 
by a little amount from the probability of the pure system (i.e.,
$h_i(t)=0$). Increasing the amplitude to $A=0.02$ (Fig. \ref{comp_mag}
(b)), reduction of $M(t)$s increases. 
Furthermore we find that $M(t)$ for the local magnetic structure 
remains larger than that of a free spin. 
The errorbar shows the first standard deviation of the distribution of $\{M(t)\}$.
Here we find that the distribution is rather wide but the mean of the distribution 
is definitely different.
The reduction of $M(t)$ by field fluctuation corresponds to the reduction
of $p$ in Eq. (\ref{prob}).
The reduction of $p$  has already pointed out for a single spin
by Y. Kayanuma and H. Nakayama~\cite{Kaya1}. 
Our results are qualitatively consistent with their results. 
Furthermore, the present observation indicates that a local magnetic structure induced by an
impurity is less sensitive to the noise. 
This feature may be due to this bulky structure of the 
local magnetic structure. From this observation we may expect that local
magnetizations are easier to manipulate than single spins in a field with
noise. 
Detail analysis for the noise dependence will be reported elsewhere.

\section{Summary and Discussion}
\label{sec:summary}

In the bond-alternating ($S=1/2$) Heisenberg antiferromagnetic chain,
we studied properties of 
locally induced magnetic structures by inhomogeneities of the lattice
such as the defect of alternation or the edge of the lattice.
Because of the gapful nature, the role of the inhomogeneities is
similar to that in the Haldane systems and is very different from that in the
uniform chain.  
In the uniform chain we found that the bond impurities divide the system into 
domains.
We showed that a local magnetic structure induced by a bond impurity can be
looked on as an effective $S=1/2$ spin. 
The interaction between the local magnetic moments 
decays exponentially in the present model and 
the local magnetic structures behave
almost independently even at very low temperatures.

We also studied the force acting between these defects of alternation. 
There are several configurations of defects. We investigated the forces between $J_1J_1$ and $J_1J_1$, between $J_2J_2$ 
and $J_2J_2$, and between $J_1J_1$ and $J_2J_2$.
Furthermore, there are two types of separations of defects, i.e.,
the type of (\ref{ichi}) and (\ref{stand}).  
It turned out that the forces are attractive for all cases.

We also studied the dynamical response of the magnetization to 
a sweeping field. The induced local moment behaves as a single spin
even in dynamical property as far as the uniform field is applied.   
We considered what property is different between the local magnetic
structure (effective $S=1/2$ spin) induced by an impurity and a free
$S=1/2$ spin.  In realistic situation there exists noise which is not
necessarily uniform but different from site to site.
It has been found that even when noise is applied individually at each site,
the total effect of the noise on the
broadening of the energy levels is almost the same as that of the 
single spin.
However, the dynamical response of the magnetization to a
time-dependent magnetic field with a random noise is found to be different
from that of a single spin.
That is, the local magnetic structure is found to be more robust against noise 
applied at each site individually.

The present study is the first attempt treating the dynamics of such a local
magnetic structure. We hope that this work 
provides a basic information for the manipulation 
of microscopic or nanoscale magnetic devices in the future. 
We also hope that the present study helps to analyze the magnetic properties 
at low temperatures such as observed by NMR measurement.

\acknowledgements

The present authors would like to thank Professor Jean-Paul Boucher
for his valuable and encouraging discussion.
The present work was supported by Grant-in-Aid for Scientific
Research from Ministry of Education, Science, Sports and
Culture of Japan.
M. N. was also supported by the Research Fellowships of the Japan 
Society for the Promotion of Science for Young Scientists.

\appendix

\section{The correlation length of bond-alternating chain ($S=1/2$)}
\label{app}

We determined the correlation length of HAF bond-alternating chain
($S=1/2$) in the ground state as a function of the ratio $J_1/J_2$.  
We calculated the correlation function for various ratios ($J_1/J_2$) and
estimated the correlation length using the relation 
\begin{equation}
{\rm ln} \langle S_i^z S_j^z \rangle = {\rm constant} - \frac{r_{ij}}{\xi},
\end{equation} 
where $r_{ij}$ denotes the distance between site $i$ and site $j$, and
$\xi$ denotes the correlation length of the system.
We treated periodic chains with 60 sites and 
simulations were performed at $T=0.01$ in the $M_z=0$ space. 
We show the correlation lengths for various ratios ($J_1/J_2$) in Table
\ref{corr_length}. Here we adopt the length of a pair of bonds $J_1$
and $J_2$ to be a unit length, 
i.e., this unit is double of the bond length.

\begin{figure}
\caption{
Magnetization profiles $\{m_i\}$ of the models (a)-(d) at $T=0.01$ in the
$M_{z}=1/2$ space.
(a) and (c) contain 63 sites ($L=63$) and (b) and (d) contain 65 sites ($L=65$) .
The diamonds denote the strength of bonds $\{J_{i,i+1}\}$, those at high
positions 
denote $J_1=1.3$ and those at the low positions denote $J_2=0.7$.
Details are shown in text.
}
\label{fig_bondalt_mag}
\end{figure}

\begin{figure}
\caption{
Summation of magnetization per site from the
left edge site for the model (b).
}
\label{fig_bondalt_sum}
\end{figure}

\begin{figure}
\caption{Summation  of magnetization per site from the left edge site for a bond-alternating chain of the type (b) in Fig. \ref{fig_bondalt_mag} with $L=21$.
The strong and weak bonds are $J_{1}=2$ and $J_{2}=0.5$, respectively.}
\label{fig_bondalt_diagmag}
\end{figure}

\begin{figure}
\caption{Distribution of $M_{z}$  for the model (b) in Fig.
\ref{fig_bondalt_mag}.
}
\label{fig_bondalt_prob}
\end{figure}

\begin{figure}
\caption{Magnetization profile $\{m_i\}$ for a model which has five
positions of induced 
moments in the $M_z=1/2$ space at $T=0.01$ with $L=101$.
The diamonds denote the strength of bonds $\{J_{i,i+1}\}$, those at high
positions 
denote $J_1=1.3$ and those at the low positions denote $J_2=0.7$.}
\label{five_moment}
\end{figure}

\begin{figure}
\caption{
Ground state energy as a function
of (a) $\Delta_a$ and (b) $\Delta_b$ obtained by an exact diagonalization ($L=24$, PBC).
}
\label{fig_force_alt_diag}
\end{figure}

\begin{figure}
\caption{Avoided level crossing.
}
\label{level_cross}
\end{figure}

\begin{figure}
\caption{
Time evolution of magnetization.
Thin dotted line denotes the magnetization without noise.
Thick line denotes the magnetization of a local magnetic structure 
(effective $S=1/2$ spin) of $L=5$ with a random noise (averaged over 500 samples 
$\{h_i\}$).
Thick dotted line denotes the magnetization of a free $S=1/2$ 
spin with a random noise (averaged over 500 samples 
$\{h_i\}$). 
(a) $A=0.01, \gamma=0.1$, (b) $A=0.02, \gamma=0.1$.
}
\label{comp_mag}
\end{figure}

\begin{figure}
\caption{Energy structure of the two level system with a random 
noise for $A=0.02, \gamma=0.1$.
}
\label{level_noise}
\end{figure}

\newpage
\begin{table}
\caption{Average energy gaps $\langle \Delta E \rangle$ and their widths
$\delta E$ for several values of $A/\sqrt{2\gamma}$.
(1), (5), and (9) mean a single free spin, the minimum model consisting of
5 spins, and the next minimum model consisting of 9 spins of the type (d)
in Fig. \ref{fig_bondalt_mag}, 
respectively.
The subscript p denotes that the quantities are obtained by the
perturbation method and the subscript d denotes that the quantities are
obtained by the exact diagonalization method.
}
\label{Delta_E}
\end{table}

\begin{table}
\caption{Correlation length $\xi$ as a function of the ratio of the 
$J_1/J_2$ in the bond-alternating chain ($S=1/2$)}
\label{corr_length}
\end{table}

\end{document}